\documentclass[letterpaper,prb]{revtex4}

 \usepackage{graphicx}
 \usepackage{dcolumn}
 \usepackage{amsmath}
 \usepackage{float}
 \DeclareMathAlphabet{\mathds}{OT1}{pzc}{m}{it}

\begin{document}

 \title{Proposal for Chiral Detection by the AC Stark-Effect}
 \author{Kevin K. Lehmann}
 \affiliation{
 Departments of Chemistry \& Physics, University of Virginia, Charlottesville VA, 22904-4319}
 \email{lehmann@virginia.edu}
 \date{\today}

\begin{abstract}
 Recently, two related three wave mixing experiments have been demonstrated that use allowed rotational transitions to produce a free induction decay signal with an amplitude linearly proportional to the enantiomeric excess of a chiral molecule.   In the present work, a formally five wave mixing experiment is proposed that will exploit near-resonant AC Stark shifts to differentially split a rotational transition of R and S versions of a molecule and thus will allow for the separate measurement of the densities of both enantiomers.   
 \end{abstract}
 
 \maketitle
 

Molecular Chirality has long been a subject of intense interest to chemists for both esthetic and practical reasons.~\cite{Mislow02}   The homochirality of the principle biological components (sugars, amino acids, and nucleic acids) is fundamental to the ability to grow large polymers of defined structures and to much of molecular recognition.~\cite{Johnson05}   Chiral molecules come in enantiomeric pairs (often designated $R$ and $S$)  that can be mapped into one another by a reflection in a plane.  Sans tiny parity violating interactions, a pair of enantiomers have identical thermodynamic properties, and most physical and spectroscopic ones as well.  One-well known exception is that they interact differently with circular polarized light (which itself has a handedness and thus is chiral) leading to optical rotation and circular dichroism which are proportional to the differences in the respectively real and imaginary dielectric constants of chiral samples.   These difference arises from an interference of contributions of electric and magnetic dipole components of transitions.  Since magnetic dipole matrix elements are typically much smaller than electric dipole ones, these effects are weak and difficult to observe in the gas phase.  As one goes to longer wavelengths (corresponding to slower motions), allowed magnetic dipole transition moments become even smaller relative to electric dipole transition moments, and as a result, more difficult to observe.   In particular, optical activity has not been observed in the rotational spectroscopy of molecules.   This is unfortunate as rotational spectroscopy provide the greatest selectivity in distinguishing complex molecules of any method that can be applied in the gas phase. 

Kral \textit{et. al.}\cite{Kral03} pointed out that chiral molecules have a unique spectroscopic property.  As a consequence of their lack of definite parity, they have closed cycles of three electric dipole allowed transitions of the form $a \rightarrow b \rightarrow c \rightarrow a$.   Furthermore, the product of the three transition dipole moments will be of opposite sign for a pair of enantiomers.   Following publications of the same group\cite{Kral05,Gerbasi06} examined use of such cycles to achieve enantiomeric selection, though they explicitly assumed that two of the levels were such that they could be accessed by both L and R forms of the molecule, which requires that they be in a state with no or a low barrier to isomerization.   Hirota\cite{Hirota12} exploited the same principle of existence of cycles of three transitions but for excitation of Chiral, $C_1$ symmetry asymmetric top molecules in their ground state.   In this case, the three allowed transition moments are proportional to the projection of the permanent dipole moment (in the molecular frame) projected onto a different inertial axis of the molecule.  Even though the individual transition dipole matrix elements depends upon choices for directions for molecular axis and state phase conventions, the triple product of matrix elements is invariant and of opposite sign for a pair of enantiomers.  The sign of this triple product can be determined by the interference of amplitude transferred between states $a$ and $c$ by the direct $ a \rightarrow c$ transition and the indirect $a \rightarrow b \rightarrow c$ pair of transitions.  Hirota proposed that such a scheme can be used to spectroscopically distinguish between a pair of enantiomers.   Hirota's theoretical work was followed by an experiment by Patterson, Schnell, and Doyle~\cite{Patterson13} where they demonstrated chiral sensitivity using rotational spectroscopy.  In the initial experiment, a chiral molecule was excited with pulsed $X$ polarized radiation in the presence of a $Z$ polarized DC Stark Field.  When the Stark Field was rapidly shorted to zero after the excitation pulse, the sample produced $Y$ polarized emission that was of opposite sign for the $R$ and $S$ enantiomers and zero (by interference of the emission) for a racemic  mixture.  This effect can be understood as arising from the interference of two components of the electric transition dipole, with one induced by Stark mixing.  This work was soon followed by a second paper by Patterson and Doyle,~\cite{Patterson13b} where the DC Stark field was replaced by an RF field that was tuned on resonance, producing coherent electric dipole emission at the frequency of the $a-c$ transition induced by two photon excitation of the $a \leftrightarrow b$ and $b \leftrightarrow c$ transitions.  The emitted field was polarized perpendicular to the fields driving the $ a \leftrightarrow b$ and $b \leftrightarrow c$ transitions.  Again, there was observed a $\pi$ phase shift between the emissions from the $R$ and $S$ enantiomers and thus zero emission intensity for a racemic mixture.  

Both of these experimental approaches produce a signal that is proportional to the enantiomeric excess ($ee$) of the sample and thus can be expected to produce weak signals for sample with small \textit{ee}.  In this paper, I propose a method to produce a differential splitting of $R$ and $S$ rotational transitions produced by a chiral selective AC Stark Shift.   This AC Stark shift is produced in third order by the interaction of three mutually perpendicularly polarized electric fields that drive the three transitions in a $a \rightarrow b \rightarrow c \rightarrow a$ cycle.   For a simple three state model, one can arrange the experimental conditions such that the second order AC Stark shifts cancel and only the chiral selective third order AC Stark shift remains, leading to opposite sign spectral shifts for an $R, S$ pair.  This situation is worked out and the optimal driving condition explored.   I then turn attention to a more realistic model that accounts for the $2J+1$ degeneracy of rotational states with total angular momentum quantum number $J$. It is found that one cannot ``tune out'' the second order AC Stark shift of more than one $M$ component of the transition due to the requirement that the excitation fields be perpendicularly polarized.   Despite this, simulation of the absorption of the ``dressed states'' in the presence of three fields produces substantial chiral differential shifts.   Based upon this analysis, I consider this a promising approach to chiral analysis of complex molecules.  In particular, one should be able to make accurate relative intensity measurements of the spectrally resolved signals from the $R$ and $S$ isomers and this could lead to accurate measurement of the $\textit{ee}$ of an enantio-enriched sample, perhaps even when the magnitude of the $ee$ is only slightly less than unity.

\section{Three Level Model}

Consider the energy level diagram with three states we will label $a, b, c$ in energy order with respective zero field energy values $\hbar \omega_a, \hbar \omega_b, \hbar \omega_c$ and we use the notation $\omega_{ij} = \omega_{i} - \omega_{j}$.   We apply radiation $E(t) = E_1 \cos(\omega_1 t + \phi_1) \hat{X} + E_2 \cos(\omega_2 t + \phi_2) \hat{Y} + E_3 \cos(\omega_3 t + \phi_3) \hat{Z}$ with $\omega_1 \approx \omega_{ba}, \omega_2 \approx \omega_{cb}$, and $\omega_{3} = \omega_1 + \omega_2 \approx \omega_{ca}$.  $\hat{X}, \hat{Y}, \hat{Z}$ are three orthogonal unit vectors forming a right handed coordinate system.   We define Rabi frequencies as $\Omega_{ba} = \Omega_{ab}^* =  \left< b | \mu_X | a \right> E_1 \exp(- i \phi_1 ) / 2\hbar , \Omega_{cb} = \Omega_{bc}^* = \left< c | \mu_Y | b \right> E_2 \exp( - i \phi_2 ) / 2\hbar$, and $\Omega_{ca} =  \Omega_{ac}^* = \left< c | \mu_Z | a \right> E_1 \exp( -i \phi_3 ) / 2\hbar$.   We also define detunings $\Delta \omega_b = \omega_{ba} - \omega_1 $ and $\Delta \omega_c = \omega_{ca} - \omega_3$.   The detuning of the $b \rightarrow c$ transition is $\Delta \omega_c - \Delta \omega_b$.   We write the time dependent wavefunction in the Hilbert spaced spanned by these three states as:
\begin{equation}
\Psi(t) = C_a \, e^{-i \omega_a t } | a > + \, C_b \, e^{-i (\omega_a + \omega_1) t} |b> + \, C_c \, e^{-i (\omega_a + \omega_3) t} |c> 
\end{equation}
The time dependent Schr{\"o}dinger equation, after making the ``rotating wave'' approximation of neglecting far-off resonance terms, gives for the time dependence for the amplitudes:
\begin{equation}
i \hbar \frac{d}{dt} 
\left(
\begin{array}{c}
 C_a   \\
 C_b   \\
  C_c   
\end{array}
\right)  =  \hbar 
\left(
\begin{array}{ccc}
  0 & \Omega_{ab}  & \Omega_{ac}  \\
 \Omega_{ab}^*   & \Delta \omega_b & \Omega_{bc}   \\
\Omega_{ac}^*   &  \Omega_{bc}^*  & \Delta \omega_c   
\end{array}
\right)
\left(
\begin{array}{c}
 C_a   \\
 C_b   \\
  C_c   
\end{array}
\right) = H_{ds} \left(
\begin{array}{c}
 C_a   \\
 C_b   \\
  C_c   
\end{array}
\right)
\end{equation}
which is the same as for the wavefunction coefficients with a time independent ``dressed state''~\cite{Shirley65} Hamiltonian matrix, $H_{ds}$.    Note that if we had
not imposed the constraint that $\omega_3 = \omega_1 + \omega_2$, we would not have eliminated all the explicit time dependent terms in $H_{ds}$.

If we apply third order perturbation theory to the eigenstates of $H_{ds}$ (and correct for the energy shifts implied by the explicit time dependence in the definition of the $C_k$'s), we get AC Stark shifts of the energy levels:
\begin{eqnarray}
 E_a   /  \hbar& = & \omega_a - \frac{ |\Omega_{ab}|^2 }{ \Delta \omega_b } -  \frac{ |\Omega_{ac}|^2 }{ \Delta \omega_c }
+  \frac{ 2 \Re \left( \Omega_{ac} \Omega_{cb} \Omega_{ba} \right)    }{ \Delta \omega_b   \, \Delta \omega_c    } \\
E_b /  \hbar & = & \omega_b + \frac{ |\Omega_{ab}|^2 }{ \Delta \omega_b } +  \frac{ |\Omega_{bc}|^2 }{ \Delta \omega_b - \Delta \omega_c }
+  \frac{ 2 \Re \left( \Omega_{ba} \Omega_{ac} \Omega_{cb} \right)    }{ \Delta \omega_b   \, (\Delta \omega_b -\Delta \omega_c  )  } \\
E_c /  \hbar & = & \omega_c + \frac{ |\Omega_{ac}|^2 }{ \Delta \omega_c } -  \frac{ |\Omega_{bc}|^2 }{ \Delta \omega_b - \Delta \omega_c }
+  \frac{ 2 \Re \left( \Omega_{ca} \Omega_{ab} \Omega_{bc} \right)    }{ \Delta \omega_c   \, (\Delta \omega_c -\Delta \omega_b  )  }
\end{eqnarray}
where $\Re(x)$ is the Real part of $x$.  The last term in each case comes in third order and the Real apart arises from the two time reversed paths through the intermediate states, i.e. for state $a$ we can go $a \rightarrow b \rightarrow c \rightarrow a$ and $a \rightarrow c \rightarrow b \rightarrow a$.  These give the same denominators but numerators that are complex conjugates of each other.   If one observes a transition from any of the three levels $a,b,c$ to a forth level, $d$, one will find this transition shifted by the AC Stark effect of the initial state, assuming that the nonresonant Stark Shifts of state $d$ can be neglected.

Except for a  chiral molecule of $C_1$ symmetry (i.e. no symmetry elements except the identity operation), we cannot have a closed loop of three electric dipole allowed transitions such as $a \rightarrow b \rightarrow c \rightarrow a$.  This is due to the fact that each symmetry element besides the identity implies that at least one component of the electric dipole projected onto the inertial axes must be zero.  For a pair of enantiomers, the product $ \Omega_{ac} \Omega_{cb} \Omega_{ba} $  will be of opposite sign for the same sets of transitions.  For the case where the three coupled transitions are all rotational transition in a given vibronic state, each transition will have a sign determined by the sign of one of the components of the permanent electrical dipole moment projection onto a different principle axis of the inertial tensor.  
In assigning the inertial axes to the molecule, one is free to pick the sign of two of the components of the permanent dipole moment to be positive, but then the sign of the third component will be fixed by the requirement that one has a right handed coordinate system.  Thus, if we have a mixture of a pair of enantiomers, there will be a different sign of the third order AC Stark shifts of the energy levels while they are driven by the fields with the properties we invoked. The AC Stark effect proposed here, will produce a pair of peaks that are shifted in equal and opposite directions from the positions of the peaks without the closed cycle.  For example, if one is observing the frequency of the $a \leftrightarrow d$ transition, then the third order Stark shift of state $a$ can be observed by a shift in this resonance frequency when the field at $\omega_2$ is switched on.   Note that $\omega_2$  is far from resonance with any transition coming directly from state $a$.

The second order Stark terms can be expected to be larger than the third for the case where perturbation theory is reasonably accurate.   However, by selecting the signs of the two detunings to the two virtual levels from a state of interest, one can, for this simple three state model, arrange to cancel the second order shifts.  For example, if we select detunings and field amplitudes such that $|\Omega_{ab}|^2 \Delta \omega_c = - | \Omega_{ac} |^2 \Delta \omega_b$, the net second order AC Stark shift of state $a$ will be zero, leaving only the enantiometric distinguishing third order terms.  As will be demonstrated below, this does not hold when the $(2J+1)$ spatial degeneracy of states is considered.

Going to 5'th order in perturbation theory, we can write the energy of dressed state $a$ as:

\begin{eqnarray}
 E_a / \hbar &=&  \omega_a - \left( \frac{ |\Omega_{ab}|^2 }{ \Delta \omega_b } + \frac{ |\Omega_{ac}|^2 }{ \Delta \omega_c } \right)
\left( 1 + \frac{ |\Omega_{bc}|^2     }{  \Delta \omega_b \, \Delta \omega_c   }         \right) \nonumber \\
& & +  \left(  \frac{ 2 \Re \left( \Omega_{ac} \Omega_{cb} \Omega_{ba} \right)    }{ \Delta \omega_b   \, \Delta \omega_c    } \right)
\left( 1 + \frac{ | \Omega_{cb} |^2  }{ \Delta  \omega_b \, \Delta \omega_c    }     
- \left( \frac{1}{ \Delta \omega_b } + \frac{ 1}{  \Delta \omega_c }  \right)
\left( \frac{ |\Omega_{ab}|^2 }{ \Delta \omega_b } + \frac{ |\Omega_{ac}|^2 }{ \Delta \omega_c } \right)
 \right) 
\end{eqnarray}

\begin{figure}[h]
\begin{center}
\includegraphics[width=10cm]{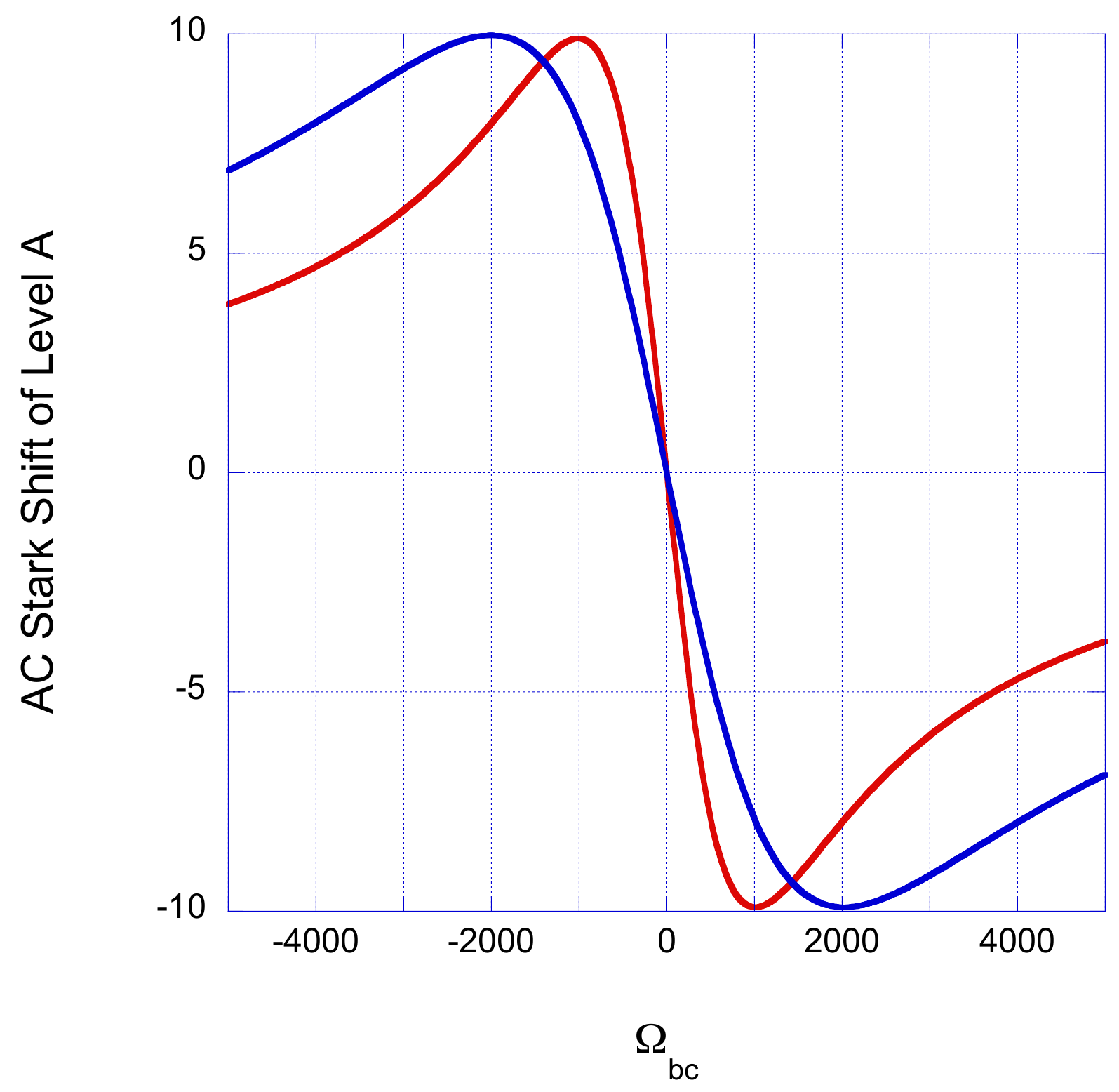}  
\caption{Plot of AC Stark shift of level $a$ as a function of $\Omega_{bc}$ for $\Omega_{ab} = 100, \Delta \omega_b = 1000$ and the cases (a) $\Omega_{ac} = 100, \Delta \omega_c = -1000$ and (b) $\Omega_{ac} = 200, \Delta \omega_c = -4000$.   Case (a) is represented by the red (more steeply rising) curve and case (b) by the blue (more slowly rising) curve. }
\label{AC3LevelPlot}
\end{center}
\end{figure}

Thus, the 4'th order terms are the same for the two enantiomers and that selecting  $|\Omega_{ab}|^2 \Delta \omega_c = - | \Omega_{ac} |^2 \Delta \omega_b$, cancels both the 2'nd and 4'th order terms.  The 5'th order terms are, like the 3'rd order terms, of opposite sign for a pair of enantiomers.  If we select parameters to cancel out the 2'nd order term, then $\Delta \omega_b \, \Delta \omega_c < 0$ and the 5'th order term will cancel part of the  3'rd order contribution.   This suggests that the optimal size of $| \Omega_{bc} |^2 \sim | \Delta \omega_b \, \Delta \omega_c |$.   This I have confirmed with numerical calculations with the $3 \times 3 \, H_{ds}$.  Figure \ref{AC3LevelPlot} shows the AC Stark shifts of level $a$ as a function of $\Omega_{bc}$  for $\Delta \omega_b = 1000, \Omega_{ab} = 100$.  The two curves are for $\Delta \omega_{c} = -1000, \Omega_{ac} = 100$ and $\Delta \omega_{c} = -4000, \Omega_{ac} = 200$, selected such that the second order AC Stark shift is zero.  In both cases, the range of AC Stark shift (as a function of $\Omega_{bc}$) is $\pm 10$, which is the same as the individual contributions to the second order Stark shifts from levels $b$ and $c$.  In both cases, the maximum shifts occur when $| \Omega_{bc} | = \sqrt{ | \Delta \omega_b  \Delta \omega_c | } $.

\section{Effect of Spatial Degeneracy}

Transitions in gas phase molecules have degeneracies that arise from the rotational symmetry of space.\cite{Zare}   These can be distinguished by specifying the projection of the total angular moment, with $M$ quantum number, on some axis, usually the $Z$ axis.   Like for the static Stark effect, this degeneracy will be at least partially lifted by the  second and  third order, chiral AC Stark effects.   In general, to account for this effect, we must use degenerate perturbation theory and diagonalize the $(2J+1) \times (2J+1)$ matrix of $M, M'$ values for the level with total rotational angular momentum quantum number $J$.  If we probe the spectrum of a transition $J \leftrightarrow J_f$ where we can neglect the AC Stark effect in state $J_f$, the AC Stark shift and splitting pattern of state $J$ will be directly observed in the spectrum.   We can expand the asymmetric top states in terms of symmetric top wavefunctions using:~\cite{Townes55}
\begin{equation}
\left|J,\tau, M \right>  = \sum_{K = -J}^{J}  \left|J, K, M \right> \, A_{K,\tau}
\end{equation}
where $\tau = 0 \dots 2J$ labels the different rotational energy levels for fixed $J$ in increasing energy order.  The more commonly used asymmetric rotor rotational state labels $K_a$ and $K_b$ can be determined using $K_a =$floor$((\tau+1)/2)$ and $K_c = $floor$((2J - \tau)/2)$ where floor$(x)$ is the largest integer $\le x$. 
The transition dipole matrix element between symmetric top states $|J', K', M' >$ and $|J, K, M >$ is given by $(E_G \mu_{\alpha} /2)  \left< J, \tau, M | \phi_{\alpha, G} | J', \tau', M'\right> $, where the integral is known as a direction cosine matrix element as it is the matrix element of the dot product of a unit vector aligned with the $G = X, Y, Z$ axis of the laboratory axes system, and the unit vector along the $\alpha = A, B, C$ principle axes of the moment of inertia tensor of the molecule, in order of increasing
moment of inertia. Townes \& Schawlow~\cite{Townes55} give explicit expressions for these, using a notation $\left< J, K, M | \phi_{\alpha, G} | J', K', M'\right> = \left< J| \phi_J | J' \right> \left< J, K | \phi_{\alpha} | J', K'\right> \left< J, M | \phi_{\alpha, G} | J', M'\right>$. I will use a notation where $\sqrt{  \left< J| \phi_J | J' \right>  }$ is included in the other two factors as then each factor is $\le 1$.
Transforming to the asymmetric top eigenfunction basis gives matrix elements:
\begin{eqnarray}
\left< J, \tau, M | \phi_{\alpha, G} | J', \tau', M'\right>  =  \left<J,\tau | \phi_{\alpha} | J', \tau' \right>  \left<J,M | \phi_G |J',M'\right> \\
\left< J, \tau  | \phi_{\alpha} | J', \tau' \right>  = \sum_{ K,K' } A_{K,\tau} A_{K',\tau'} \left<J,K|\phi_{\alpha}|J',K'\right>
\end{eqnarray}
  An allowed rotational transition will have only one nonzero $\phi_{\alpha}$ transition component~\cite{Townes55} and to have a closed cycle of three transitions, 
the three principle axes directions are each represented once. 
The triple product of Rabi frequencies that contribute to the Chiral AC Stark effect can be written:
\begin{eqnarray}
\Omega_{ab} \Omega_{bc} \Omega_{ca} & =  & -e^{- i (\phi_1 + \phi_2 - \phi_3 )} \frac{ E_1 E_2  E_3  }{8 \hbar^{3}}   \mu_{A} \mu_{B} \mu_{C}  \times  
  \left<J_a,\tau_a |\phi_{\gamma}|J_b,\tau_b \right>\left<J_b,\tau_b |\phi_{\beta}|J_c,\tau_c \right>\left<J_c,\tau_c | \phi_{\alpha}|J_a,\tau_a   \right> \\
& &  \sum_{M' = M'' \pm 1} 
\left<J_a,M|\phi_{X}|J_b,M'\right>\left<J_b,M'|\phi_{Y}|J_c,M''\right>\left<J_c,M''|\phi_{Z}|J_a,M''\right>   \nonumber
\end{eqnarray}
To be nonzero, $M = M''$ or $M'' \pm 2$.
Both the products of matrix elements $  \left<J_a,\tau_a |\phi_{\gamma}|J_b,\tau_b \right>\left<J_b,\tau_b |\phi_{\beta}|J_c,\tau_c \right>\left<J_c,\tau_c | \phi_{\alpha}|J_a,\tau_a   \right>$ and $\left<J_a,M|\phi_{X}|J_b,M'\right>\left<J_b,M'|\phi_{Y}|J_c,M''\right>\left<J_c,M''|\phi_{Z}|J_a,M''\right> $ are imaginary, giving real values for the products of direction cosine matrix elements.

\begin{figure}[H]
\begin{center}
\includegraphics[width=10cm]{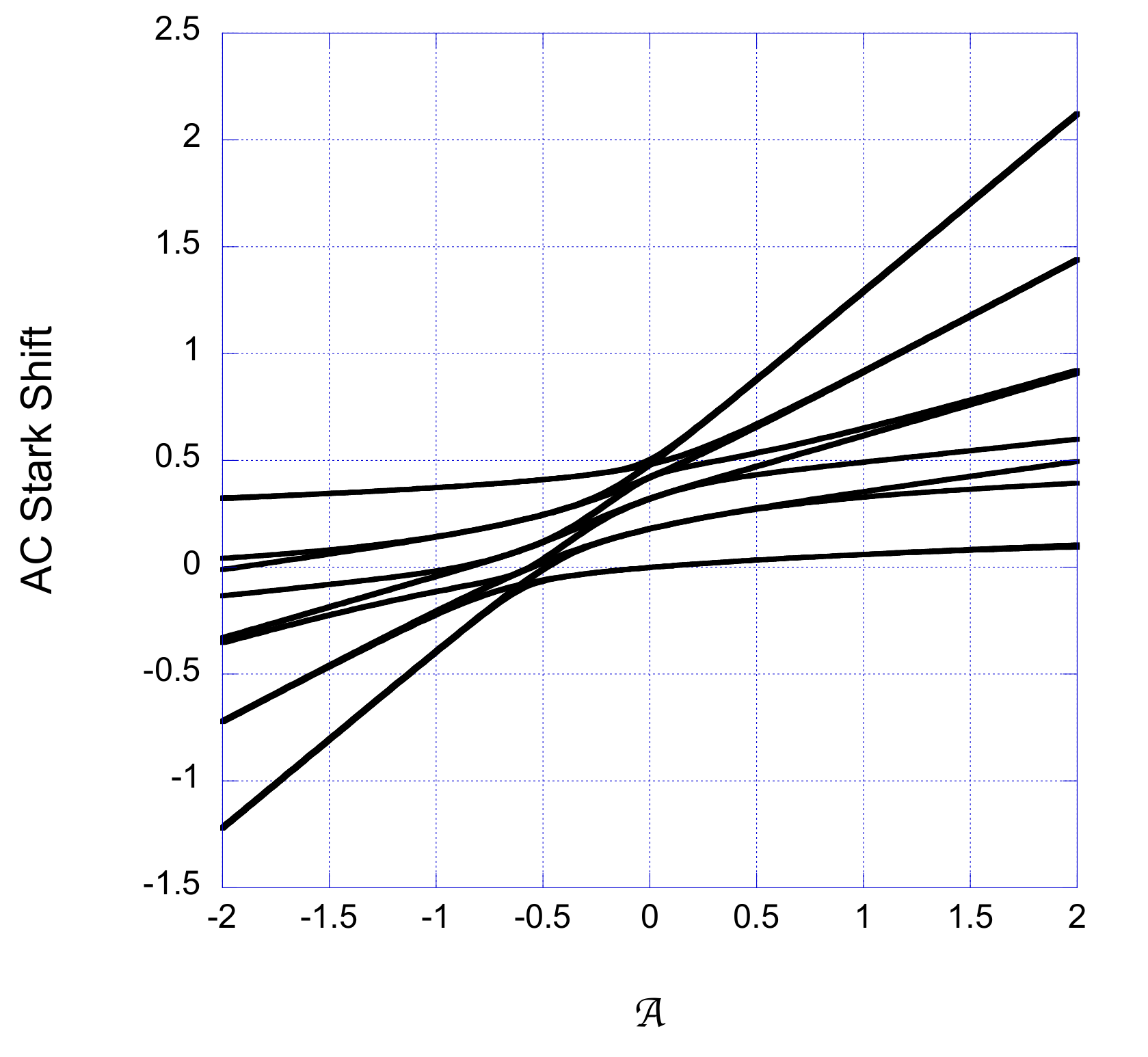}  
\caption{Plot of second order AC Stark Effect of the 11 components of a $J_a = 5$ state with two near resonant fields to states with   $J_b = 5$ and $J_c = 4$ and with ${\cal B} = 1$ (coupling between states $a$ and $b$) and variation in ${\cal A}$ from  -2 to +2. (coupling between states $a$ and $c$)}
\label{StarkXZ554}
\end{center}
\end{figure}

\begin{figure}[H]
\begin{center}
\includegraphics[width=10cm]{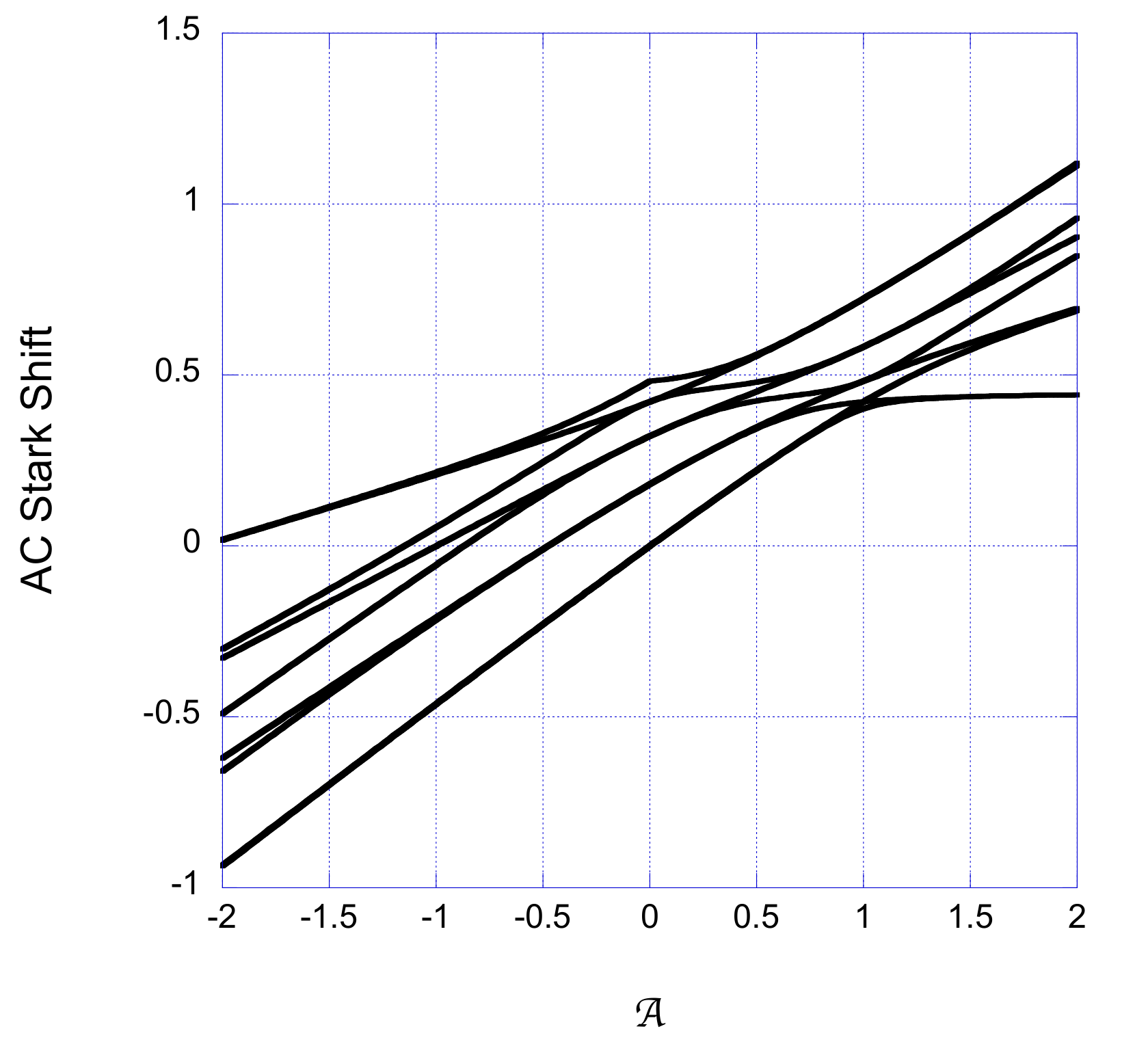}  
\caption{Plot of second order AC Stark Effect of the 11 components of a $J_a = 5$ state with two near resonant fields to states with   $J_b = 4$ and $J_c = 4$ and with ${\cal B} = 1$ (coupling between states $a$ and $b$) and variation in ${\cal A}$ from  -2 to +2. (coupling between states $a$ and $c$)}
\label{StarkXZ544}
\end{center}
\end{figure}

\begin{figure}[H]
\begin{center}
\includegraphics[width=10cm]{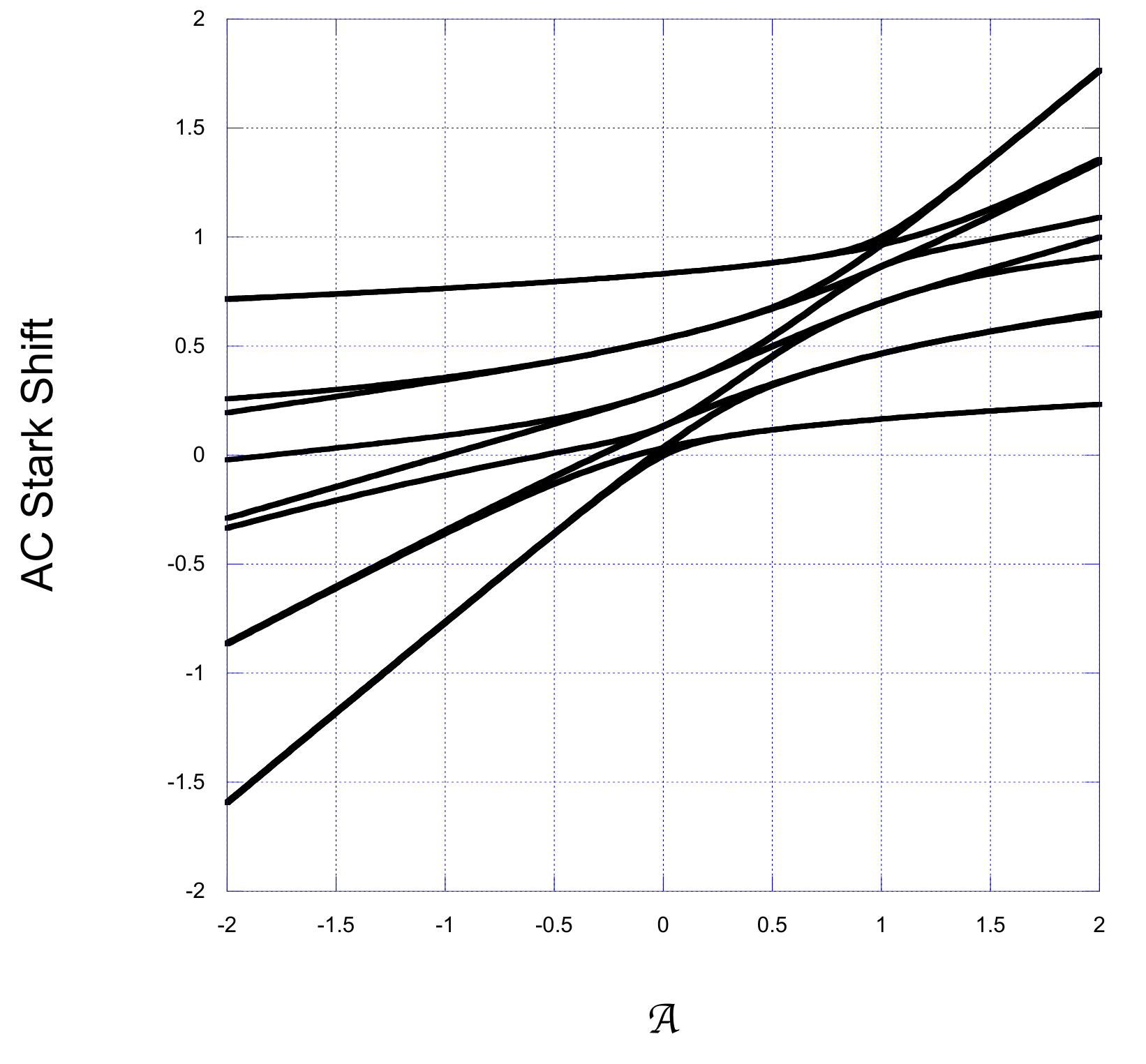}  
\caption{Plot of second order AC Stark Effect of the 11 components of a $J_a = 5$ state with two near resonant fields to states with   $J_b = 5$ and $J_c = 4$ and with ${\cal B} = 1$ (coupling between states $a$ and $b$) and variation in ${\cal A}$ from  -2 to +2. (coupling between states $a$ and $c$)}
\label{StarkXZ555}
\end{center}
\end{figure}

The effective Stark Hamiltonian for the state $J_a$ can be written with nonzero matrix elements:
\begin{eqnarray}
\left< J_a, M \pm 2 | H_{\rm eff} | J_a, M \right> &=&  {\cal B} \left< J_a, M \pm 2 | \phi_X | J_b, M \pm 1   \right> \left<  J_b, M \pm 1 | \phi_X | J_a, M   \right>    \\
& & + {\cal C} \left< J_a, M\pm 2 | \phi_X | J_b, M \pm 1   \right> \left<  J_b, M \pm 1 | \phi_Y | J_c, M    \right>  \left<  J_c, M  | \phi_Z | J_a, M   \right> \nonumber \\
& & - {\cal C} \left< J_a, M \pm 2 | \phi_Z | J_c, M \pm 2   \right> \left<  J_c, M \pm 2 | \phi_Y | J_b, M \pm 1   \right>  \left<  J_b, M \pm 1 | \phi_X | J_a, M   \right> \nonumber \\
\left< J_a, M | H_{\rm eff} | J_a, M \right> &=&   {\cal A} \left|  \left< J_a, M  | \phi_Z | J_c, M    \right>  \right|^2  +  \sum_{\pm} \left[ {\cal B} \left|  \left< J_a, M  | \phi_X | J_b, M \pm 1   \right> \right|^2  \right. \\
 && + {\cal C} \left< J_a, M | \phi_X | J_b, M \pm 1   \right> \left<  J_b, M \pm 1 | \phi_Y | J_c, M    \right>  \left<  J_c, M  | \phi_Z | J_a, M   \right> \nonumber \\
 & & -  \left.   {\cal C} \left< J_a, M  | \phi_Z | J_c, M    \right> \left<  J_c, M  | \phi_Y | J_b, M \pm 1   \right>  \left<  J_b, M \pm 1 | \phi_X | J_a, M   \right> 
\right]  \nonumber 
\end{eqnarray}
with
\begin{eqnarray}
{\cal A} &=& - \left( \frac{  \mu_{\alpha} E_3  }{ 2 \hbar   }  \right)^2 \frac{  \left|  \left<   J_a, \tau_a | \phi_{\alpha} | J_c, \tau_c     \right>  \right|^2 }{   \Delta \omega_c  } \\
{\cal B} &=& - \left( \frac{  \mu_{\beta} E_1  }{ 2 \hbar   }  \right)^2 \frac{  \left|  \left<   J_a, \tau_a | \phi_{\beta} | J_b, \tau_b     \right>  \right|^2 }{   \Delta \omega_b  } \\
{\cal C} &=& - \cos (\phi_1 + \phi_2 - \phi_3 ) \frac{ E_1 E_2  E_3  }{8 \hbar^{3}}   \mu_{A} \mu_{B} \mu_{C}  \times  
  \left<J_a,\tau_a |\phi_{\gamma}|J_b,\tau_b \right>\left<J_b,\tau_b |\phi_{\beta}|J_c,\tau_c \right>\left<J_c,\tau_c | \phi_{\alpha}|J_a,\tau_a   \right> 
\end{eqnarray}
If we take ${\cal B, C} = 0$, we have the traditional AC Stark Effect with $M$ along the axis of the field a good quantum number and Stark Shifts of 
${\cal A}$ times $((J+1)^2 - M^2)/ (J+1)\sqrt{ (2J+1)(2J+3)} $, $M^2/J(J+1)$, or $\left( J^2 - M^2 \right)/ J \sqrt{ 4 J^2 -1 }$ for $J_c = J_a + 1, J_a$, and $J_a - 1$ respectively.   If we take ${\cal A, C} = 0$, we get the same energy levels after substituting ${\cal C}$ for ${\cal A}$ and $J_b$ for $J_c$.  However, in this case the states are eigenstates of $J_X$, not $J_Z$.    If we take ${\cal C} = 0$, we have a double second order AC Stark effect.  Unfortunately, we can only partially cancel the two Stark Effects!   Figures \ref{StarkXZ554},\ref{StarkXZ544}, and \ref{StarkXZ555} show plots of the AC Stark shifts of the $M$ components of a $J_a = 5$ state coupled by near resonant fields to states with ($J_b = 5, J_c = 4$), ($J_b = 4, J_c = 4$), and ($J_b = 5, J_c = 5$) respectively.  In each case, calculations are for ${\cal B} = 1$ and ${\cal A}$ is varied from -2 to +2.  For the case where  $J_b \ne J_c $ the smallest variance of the eigenvalues for fixed ${\cal B}$ is when ${\cal A} = -0.302 {\cal B}$ for which the variance of the energy levels is reduced by a factor of 0.75 from the ${\cal A} = 0$ case.   For both the $J_b = J_c$ cases,  the smallest variance for fixed ${\cal B}$ is for ${\cal A} = 0.5 {\cal B}$ with, again, the variance 0.75 times as much as for ${\cal B} = 0$.   Thus, we can realize only a limited degree of cancelation of the second order Stark Effects, regardless the relative strengths or detunings of the two near resonant excitations.  However, note that in the $J_b = J_c$ cases,  the optimal second order shift occurs when ${\cal B}$ and ${\cal A}$ have the same signs, which implies we can take $\Delta \omega_a$ and $\Delta \omega_c$ to have the same signs, including setting them equal to one another.  If we do this, we will drive the $b \leftrightarrow c$ transition exactly on resonance, for which we can expect a much larger effect.


If we drive the $b \leftrightarrow c$  transition exactly on resonance, we cannot expect to be able to use perturbation theory, but it straightforward to simply diagonalize the $\left[ 2(J_a + J_b + J_c) + 3 \right] \times \left[ 2(J_a + J_b + J_c) + 3 \right]$ matrix that includes all the M components of all three rotational levels.  This is a generalization of the time independent $H_{ds}$ above.  I use the notation $ \Omega_{ba} = \Omega_{ab}^* =  \mu_{\alpha} \left< J_a, \tau_a | \phi_{\alpha} | J_b, \tau_b   \right> E_1 \exp(- i \phi_1 ) / 2\hbar $ and corresponding expressions for $\Omega_{ac}$ and $\Omega_{bc}$, and now the off-diagonal matrix elements between state $ | i = a, b, c , M_i >$ and $ | j  , M_j >$ is given by $\Omega_{ij} \left< i, M_i | \phi_{G} | j, M_J \right >$, where $G$ corresponds to the polarization of radiation nearly resonate with the $i \leftrightarrow j$ transition.  One of three $\Omega$, corresponding to the $B$ polarized rotational transition, will be purely imaginary.  It makes no difference to the results if we permute the factor of $i$ between $\Omega_{ab}, \Omega_{bc}$, or $\Omega_{ac}$.  The eigenstates are made up of functions with $M_a$ even (odd), $M_c$ even (odd), and $M_b$ odd (even).  

\begin{figure}[H]
\begin{center}
\includegraphics[width=10cm]{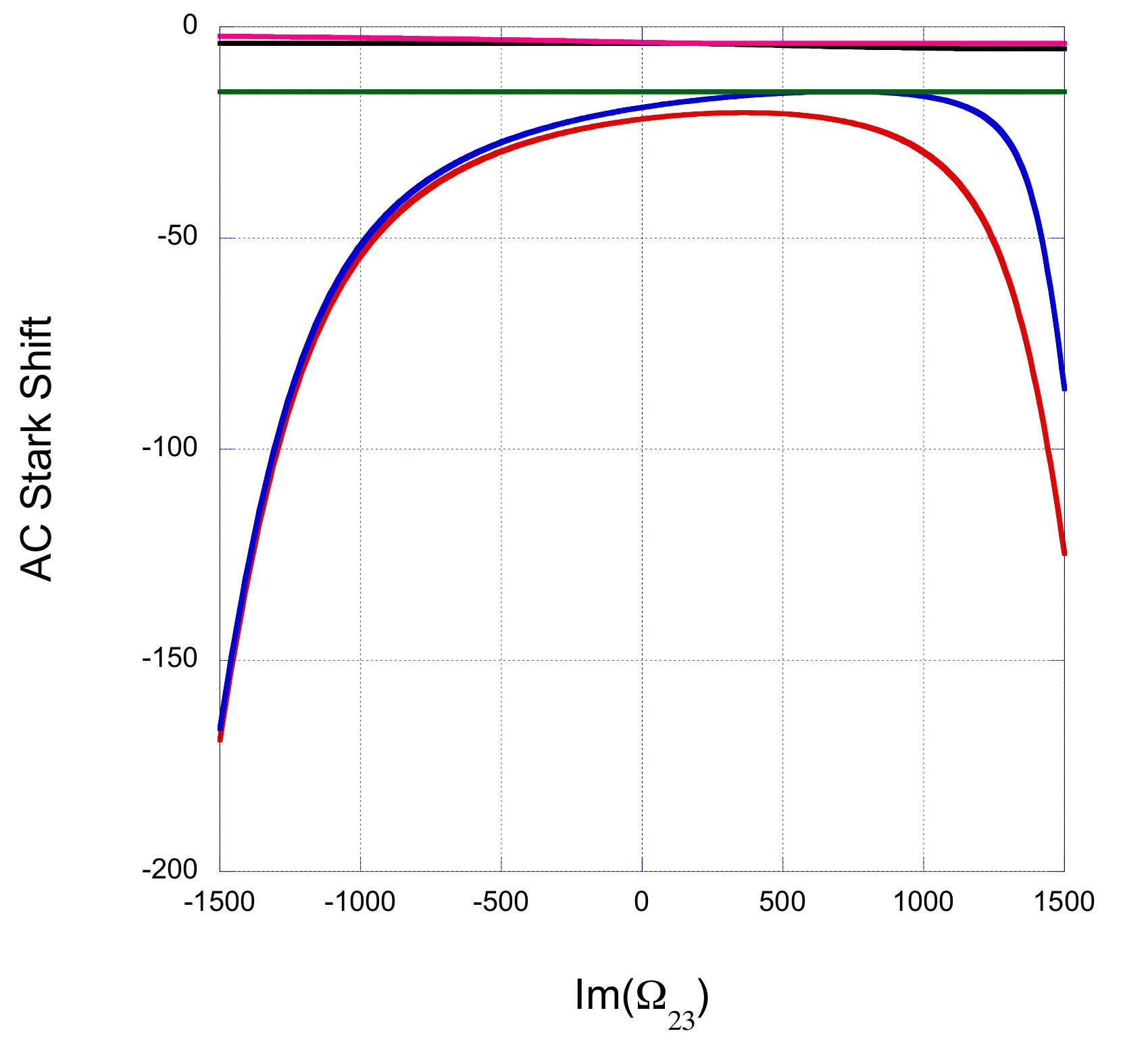}  
\caption{Plot of AC Stark Shifts of the $J_1 = 2$ state as function of $\Omega_{bc}$with $J_2 = J_3 = 1, \Omega_{ab} = 100, \Omega_{ac} = 200, \Delta \omega_b = \Delta \omega_c = 1000 $}
\label{J211_100_100}
\end{center}
\end{figure}

As an example, consider the case $J_a =2, J_c =  J_b = 1, \Delta \omega_b = \Delta \omega_c = 1000, \Omega_{ab} = \Omega_{ac} = 100$.  Figure \ref{J211_100_100} gives a plot of the five eigenvalues of states that adiabatically evolve from the $J = J_a$ eigenstates as a function of $\Omega_{bc}$ (which is assumed to imaginary).  It is evident that three of the five $J = 2$ states have AC Stark shifts that are insensitive to the value of $\Omega_{bc}$ but the lowest energy states are shifted down in frequency by an amount that depends strongly on the sign of $\Im(\Omega_{bc})$ and thus will differentially shift the transitions of a pair of enantiomers.   As $|\Omega_{bc}|$ becomes comparable to $\Delta \omega_b = \Delta \omega_c$, the chiral sensitive AC Stark shifts become large compare to the shifts when $\Omega_{bc} = 0$.  This can be understood to be a consequence of the largest resonant (and thus first order) Stark Shift of the components of the mixed $b$ and $c$ state shifting into near resonance with the dressed state energies of the components of state $a$.  For $\Omega_{bc} = -1500i$, the lowest states that arise from the two $J = 1$ states are at 104.5 and 106.5, i.e., have AC Stark shifts of 90\% of the splittings without Stark shifts.  

\begin{figure}[H]
\begin{center}
\includegraphics[width=10cm]{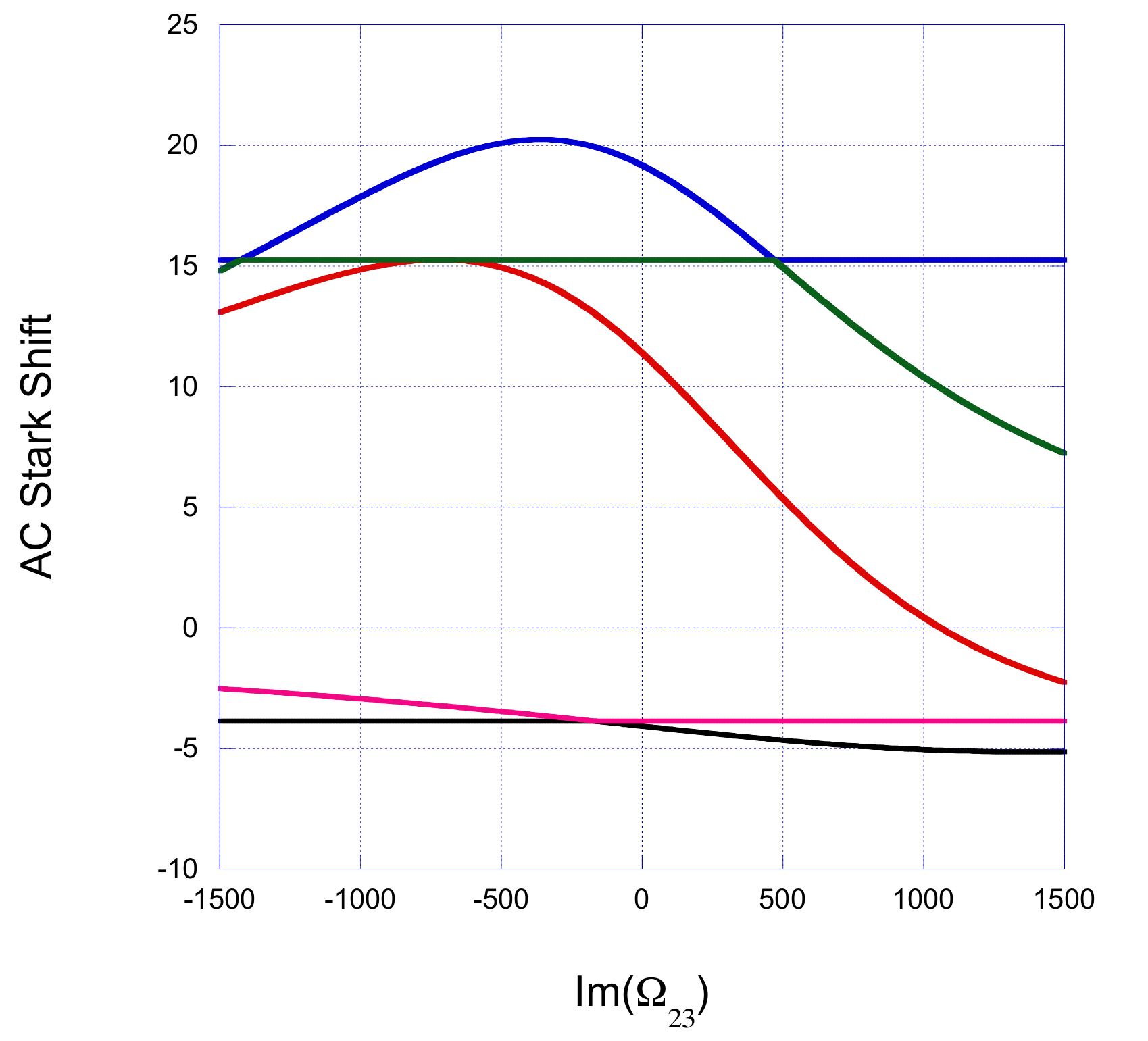}  
\caption{Plot of AC Stark Shifts of the $J_1 = 2$ state as function of $\Omega_{bc}$with $J_2 = J_3 = 1, \Omega_{ab} = 100, \Omega_{ac} = 200, \Delta \omega_b = 1000,  \Delta \omega_c = -1000 $}
\label{J211_100_M100}
\end{center}
\end{figure}

\begin{figure}[H]
\begin{center}
\includegraphics[width=10cm]{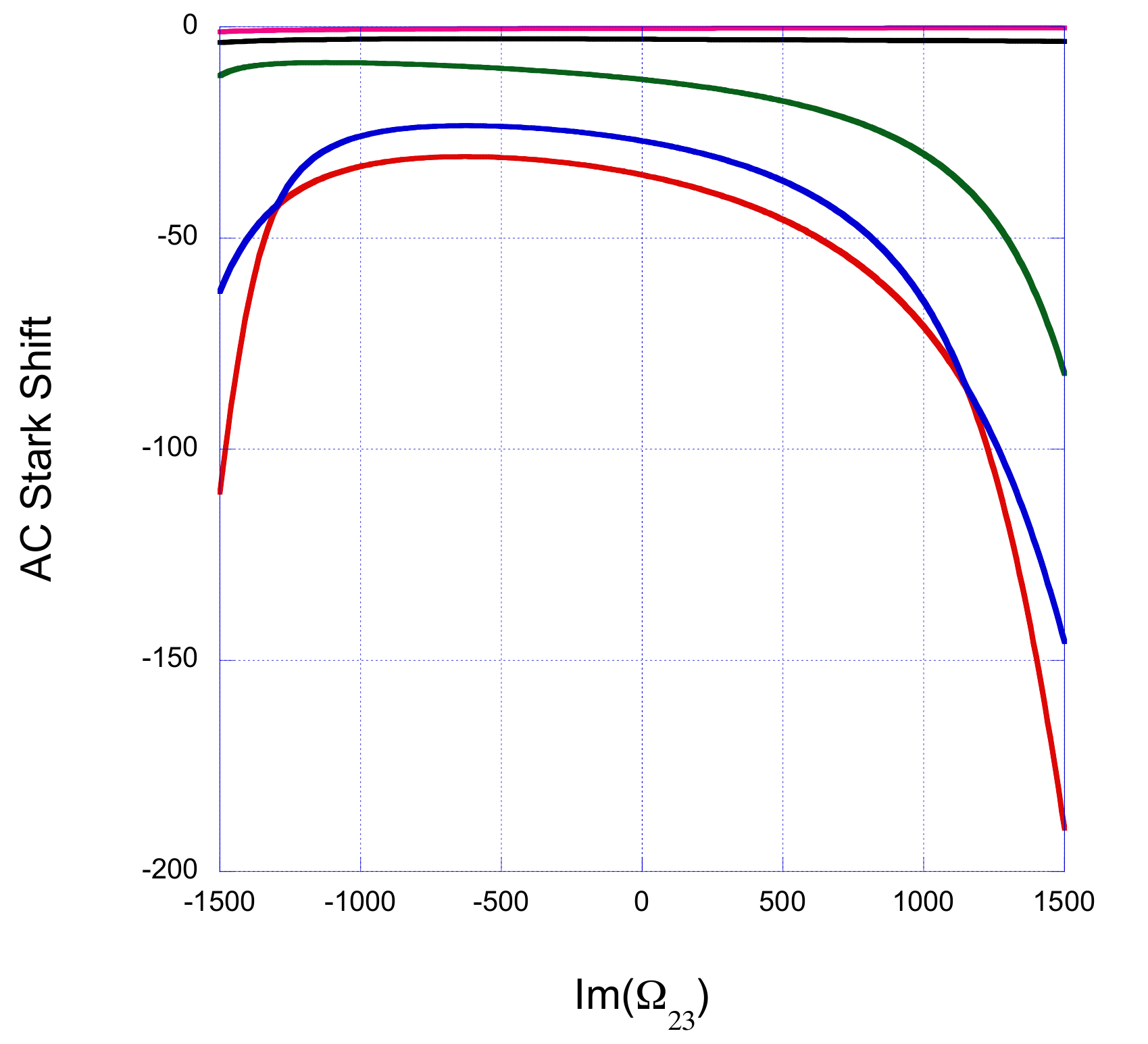}  
\caption{Plot of AC Stark Shifts of the $J_1 = 2$ state as function of $\Omega_{bc}$with $J_2 = 1,  J_3 = 2, \Omega_{ab} = 100, \Omega_{ac} = 200, \Delta \omega_b = \Delta \omega_c = 1000 $}
\label{J212_100_100}
\end{center}
\end{figure}

\begin{figure}[H]
\begin{center}
\includegraphics[width=10cm]{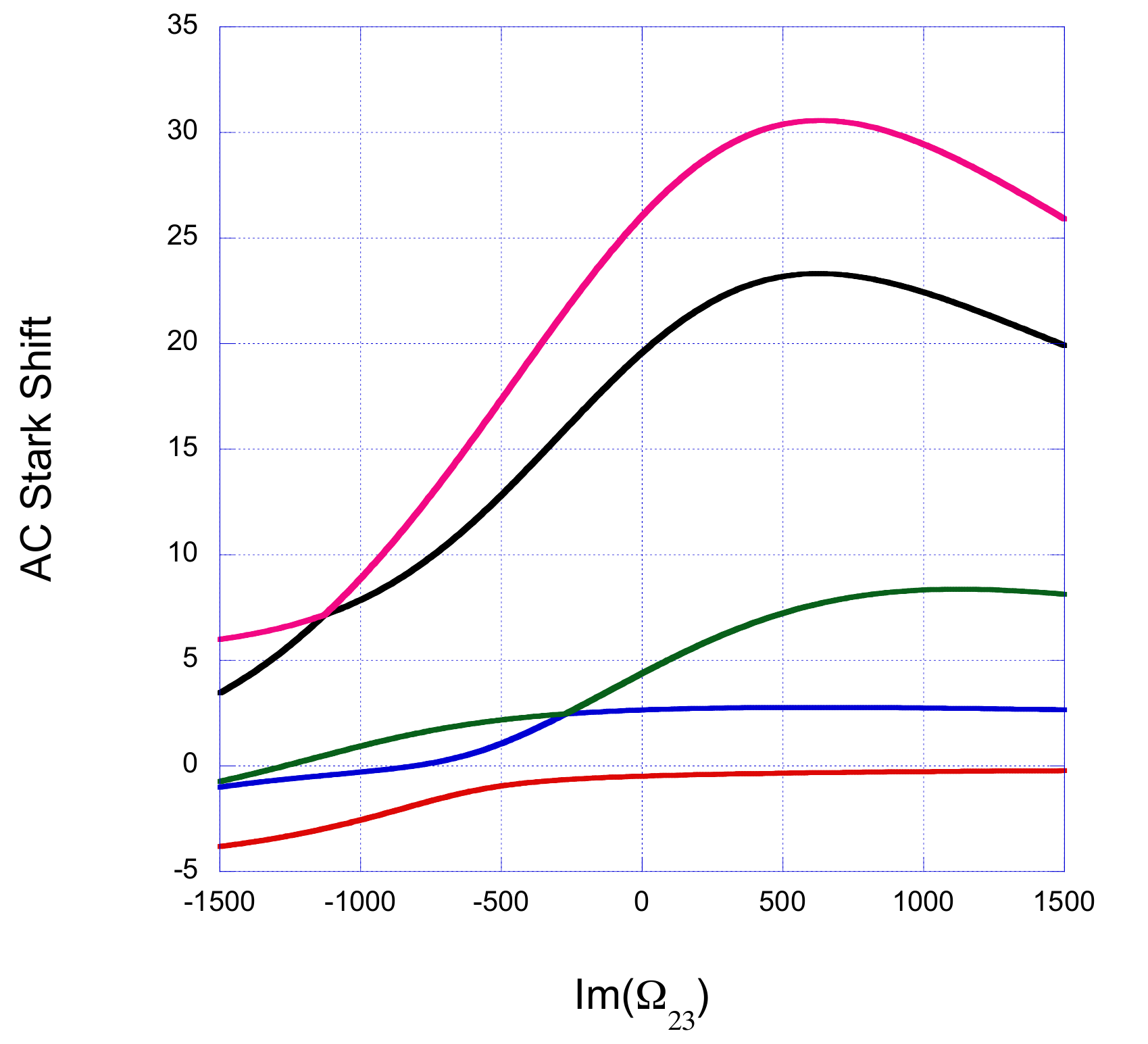}  
\caption{Plot of AC Stark Shifts of the $J_1 = 1$ state as function of $\Omega_{bc}$with $J_2 = 1,  J_3 = 2, \Omega_{ab} = 100, \Omega_{ac} = 200, \Delta \omega_b = 1000, \Delta \omega_c = -1000 $}
\label{J212_100_M100}
\end{center}
\end{figure}

If we keep everything else the same and change $\Delta \omega_b = - \Delta \omega_c$ we get the variation of dress state eigenenergies shown in Figure \ref{J211_100_M100}.  The resonance enhancement of the mixing of states $J = J_b$ and $J_c$ in evident in the much larger AC Stark Shifts in the $\Delta \omega_b = \Delta \omega_c$ case.   If we take $\Omega_{bc}$ to be real, the calculated energy values are symmetric under $\Omega_{bc} \leftrightarrow -\Omega_{bc}$, i.e. there would be no difference in the spectra of enantiomers.  Figures \ref{J212_100_100} and \ref{J212_100_M100} display the energy levels as functions of $\Omega_{bc}$ for the case with $J_a = J_c = 2, J_b = 1$ and $\Delta \omega_b = \pm \Delta \omega_c = 1000$.   Again,the resonance enhancement is evident.  In this case, three of the five states are strongly shifted, which matches the number of states of the $J_b, J_c$ resonance that are strongly shifted down towards the energy of $J_a$.

\begin{figure}[H]
\begin{center}
\includegraphics[width=10cm]{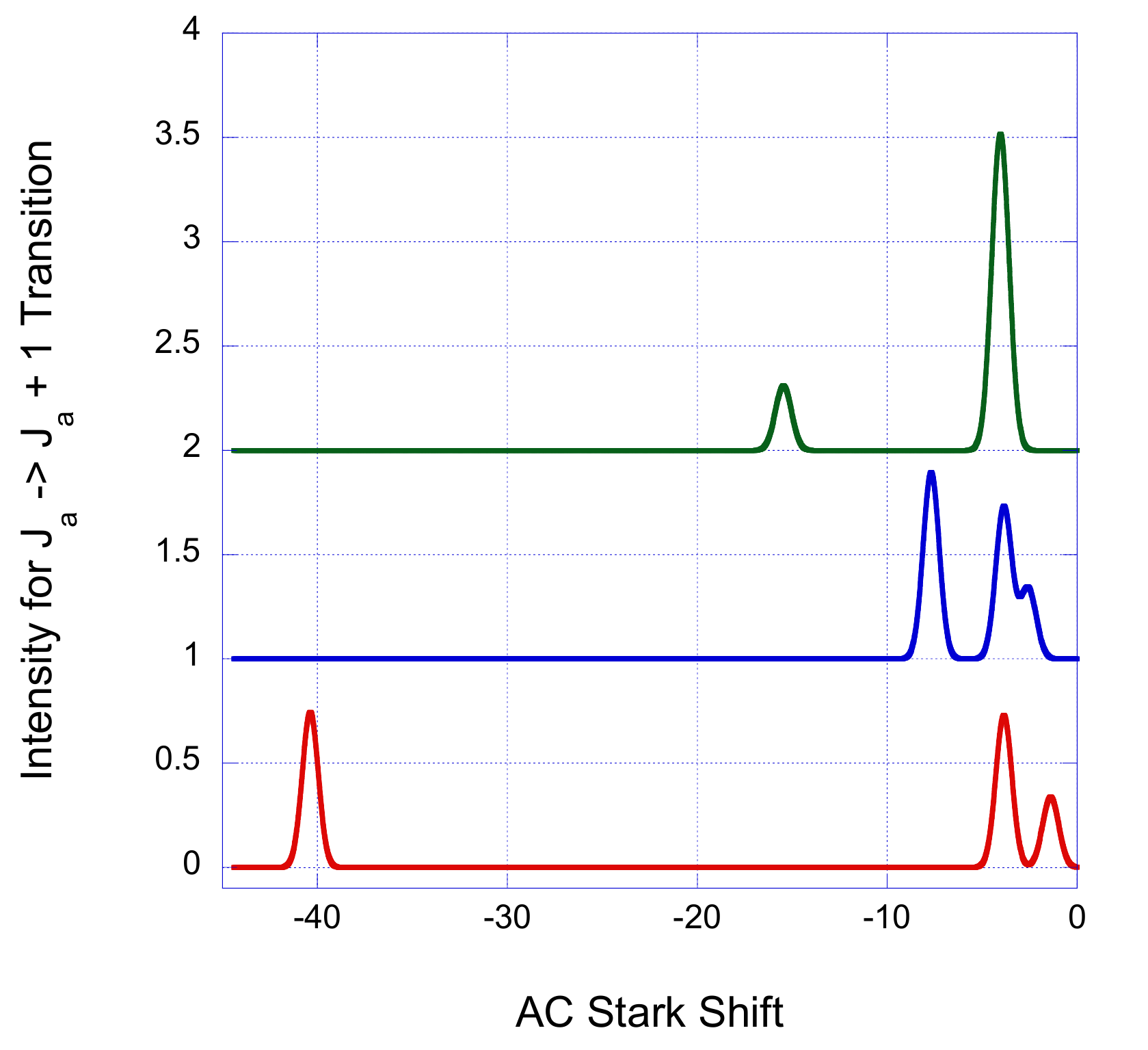}  
\caption{Simulated $J_a = 2 \leftrightarrow 3$ spectrum displaying AC Stark shifts for $J_2 = J_3 = 1, \Omega_{ab} = 100, \Omega_{ac} = 200, \Delta \omega_b = \Delta \omega_c = 1000$.  The spectra are offset for clarity, with (from bottom to top) $\Omega_{bc} = -1200i, 0, +1200i$}
\label{Spec_plot_1200}
\end{center}
\end{figure}

\begin{figure}[H]
\begin{center}
\includegraphics[width=10cm]{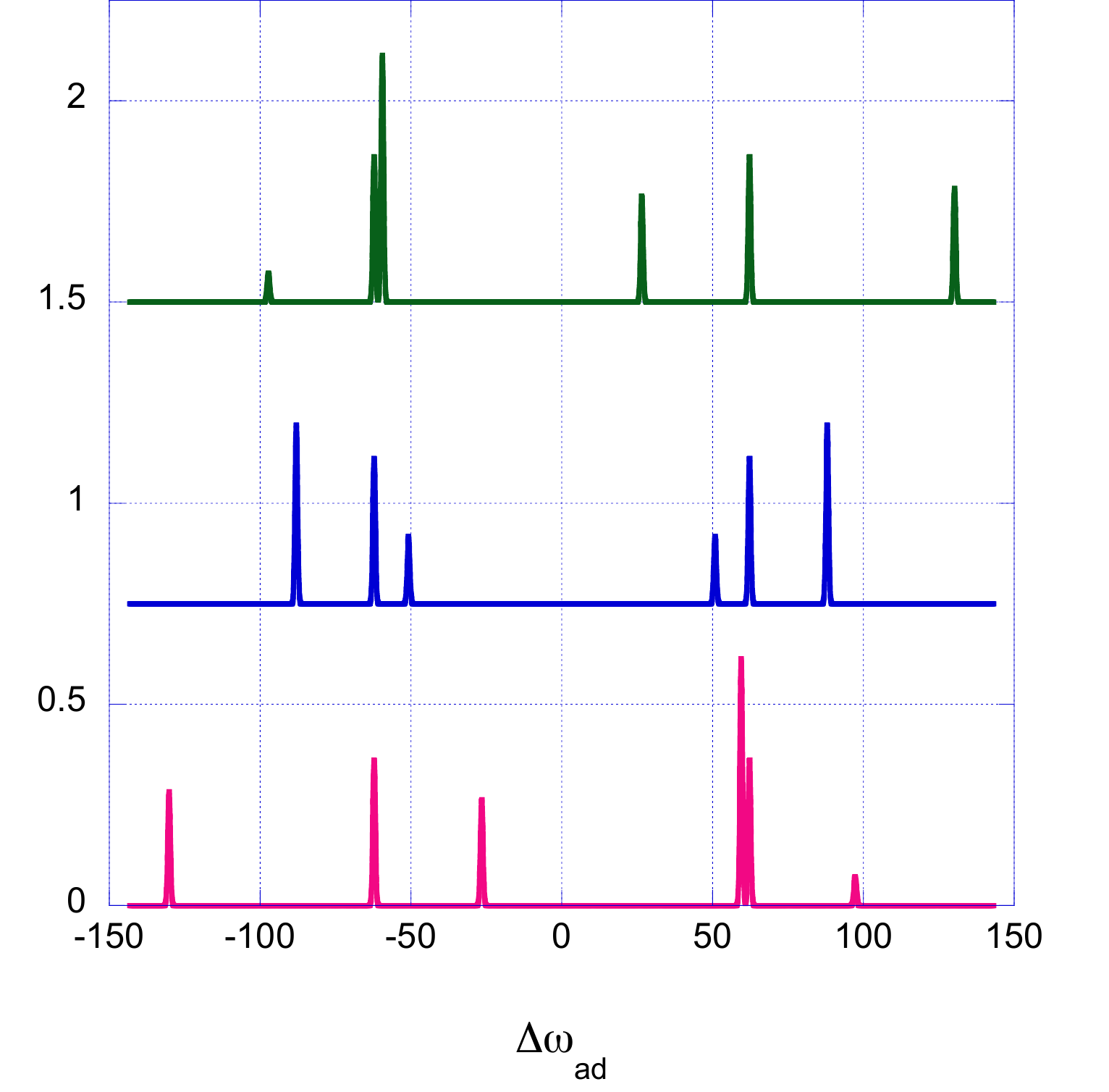}  
\caption{Simulated $J_a = 2 \leftrightarrow 3$ spectrum displaying AC Stark shifts for $J_2 = J_3 = 1, \Omega_{ab} = 100, \Omega_{ac} = 100, \Delta \omega_b = \Delta \omega_c = 0$.  The spectra are offset for clarity, with (from bottom to top) $\Omega_{bc} = -100i, 0, +100i$}
\label{spec_plot_211_100_100_100}
\end{center}
\end{figure}

The AC Stark shifts for a spectrum of a $J_a = 2 \rightarrow 3$ transition is displayed in Figure \ref{Spec_plot_1200} for $J_b = J_c = 1, \Delta \omega_b = \Delta \omega_c = 1000, \Omega_{ab} = 100, \Omega_{ac} = 200$ and for $\Omega_{bc} = -1200i, 0, +1200i$.  The plot is for an $X$ polarized probe.  All Stark components appear in the spectrum regardless of polarization of the probe field.   Figure \ref{spec_plot_211_100_100_100} displays the AC Stark shifts for the same transitions when $\Delta \omega_b = \Delta \omega_c = 0$, $\Omega_{ab} = \Omega_{ac} = 100$ and $\Omega_{bc} = -100i, 0, +100i$.  In the cases with zero detunings of the 3-cycle transitions,  the $\Delta \omega_{ad}$ spectrum is symmetric for $\Omega_{bc} = 0$.  This is lifted for $\Omega_{bc} \ne 0$ but it is symmetric for simultaneous change in the signs of $\Delta \omega_{ad}$ and $\Omega_{bc}$.

It may not be obvious to the reader why I have selected the three excitation polarizations perpendicular to one another.  If two were parallel, we get dressed state energies that are invariant to sign change of any of the $\Omega$ values and thus the same for a pair of enantiomers.   A way to rationalize this result is to note that to distinguish between the enantiomers, we need to determine if the vectors formed by the projection of the dipole moment onto three principle axes of the moment of inertia tensor form a right or left handed coordinate system.  Since interactions in the electric dipole approximation involve projections of the molecular frame dipole components onto the laboratory field components, these have to be nonplaner to distinguish between right and left handed systems.

\section{Some practical issues...}

In the discussion above, I have implicitly assumed that the magnitudes of the electric fields, $E_i$ are fixed and that the relative phases ($\phi_1+ \phi_2 - \phi_3$) are also fixed.   Any dispersion of these across the sample that is probed by the $a \leftrightarrow  d$ transition will lead to an inhomogeneous distribution of AC Stark shifts.  Because of the change of phase with propagation of the fields, we can only exactly satisfy the condition of constant relative phases if the three waves co-propagate, but this is not possible given the requirement that the three fields are all mutually perpendicular.  This is analogous to the fact that one cannot achieve phase matching in the case of the 3-wave mixing version of the experiment of Patterson and Doyle.~\cite{Patterson13b}   This effect can be minimized by taking transitions $a \rightarrow b$ and $a \rightarrow c$ to be close in frequency and co-propagating.  This means that $\omega_{bc}$ will be relatively small and can be applied as a quasi-DC field, polarized in the direction of the propagation of the other two fields.   In practice, this can be realized if the transition $b \rightarrow c$ is an asymmetry doublet of the chiral asymmetric top.

Another important consideration is if sufficient electric field amplitudes can be realized in practice.   The Rabi frequency of a transition is given by $| \Omega_{ij} | = 64 \times 10^6$\,s$^{-1}\,\sqrt{I / W cm^{-2}} ( \mu_{ij} /D )$.  Values of $\mu_{ij}$ will typically be about half the corresponding permanent dipole moment projection along the relevant axis.  If we suppose that we apply 1000 V across 10 cm gap for the low frequency $\omega_{3}$, we will have $| \Omega_{bc} | =  158 \times 10^6$\,s$^{-1} \,  ( \mu_{bc} /D )$.   The plots of AC stark shifts given above are for the angular frequencies and thus need to be reduced by $2\pi$ to put in units of Hz.  Referring to figure~\ref{Spec_plot_1200}, where the chiral differential spectral shift is a few percent of the largest Rabi frequency suggests that we can realize chiral differential shifts on the order of a few MHz.  Though complicating the experiment, it should be possible to use resonant cavities to enhance the fields.   For the Patterson and Doyle type 3-wave mixing experiment, the use of a ``regular'' cavity (without birefringence) is ruled out as one could not expect to simultaneously have cavity resonance for both the two necessarily resonant waves.  However, for the AC Stark shift, one has the waves off resonant and these detuning values can be adjusted to allow simultaneous resonance of the two driving waves in the same cavity (with perpendicular polarization).  The third driving wave for the AC Stark effect can resonated in a cavity oriented at 90 degrees to the first, so one can in principle build up all three waves.    A disadvantage of the use of microwave cavities is that one will have standing wave fields, not traveling ones, and so the problem of inhomogeneous Stark shifts will be exacerbated. 

It is interesting to compare the chiral dependent AC Stark Shift with the methods developed by Patterson \textit{et. al.}~\cite{Patterson13, Patterson13b} where one creates and then detects a net polarization in a chiral sample that is perpendicular to two other fields that are applied to the sample.  In the first published version,~\cite{Patterson13} one field is a DC Stark Field (which is switched off before detection) and the other is a resonant excitation field.  In the second,~\cite{Patterson13b} two resonant excitation fields are applied.  In both of these cases, the magnitude of the signal is proportional to the enantiomeric excess (ee) of the sample, i.e. the difference in molecular density of a pair of enantiomers.   In such an experiment, signal to noise will be maximized for samples with large $ee$'s, but it will be difficult to make an accurate measurement of the $ee$ since there is only a single signal.  While one can observe the regular Free Induction Decay of the same transition for normalization, the loss of signal strength from phase mismatching makes it difficult to use such a normalization without having a reference sample of known $ee$.  In many potential applications of these methods, one will seek to quantify a small relative concentration of one enantiomer compared to another and this would appear challenging.    For a sample with a small $ee$, the chiral signals will be small but proportional to the $ee$ and thus one can hope to make an accurate determination of $ee$.   

If one observes the chiral dependent AC Stark effect, then one can determine the $ee$ by comparing the areas of two peaks shifted from one another.  By phase reversing one the excitation fields by $\pi$, the resonant frequencies of a pair of enantiomers will interchange, so one can compare molecular absorption or emission at the same frequency to determine the ratio of the enantiomers.   Assuming that one can get good separation of the peaks, this should provide a more accurate way to determine the $ee$ of a sample, assuming it is not so large that signal from the minor enantiomer is buried in the wing of the transition from the dominant one.  For a sample with small $ee$, the AC stark measurement will give signal strengths only reduced a factor of $\sim 2$ compared to an equivalent measurement sample with a single enantiomer.  However, the determination of the $ee$ will require accurate measurement of a small change in signal strength.  Thus, the two measurement methods can be considered complimentary to some degree.  The AC Stark measurement does require the application of four distinct frequencies to to the sample, compared to two for the three wave mixing method introduced by Paterson and Doyle, i.e. it is formally a five wave mixing experiment instead of three wave mixing.  Another potential practical difficulty with the AC Stark effect detection method is that the three driving fields producing the AC Stark effect must be present when detecting the molecular absorption or free induction decay of the $a \rightarrow d$ transition.  This will require that the frequency of this transition to be sufficiently separated that one can use filters to selectively attenuate  these driving fields.   Another potential concern of the AC Stark measurement is that one will split the zero field peak into multiple $M$ components, and this will reduce the signal to noise ratio.

\section{Acknowledgments}
The author would like to thank Brooks Pate for encouragement and helpful discussions.


\begin{thebibliography}{11}
\expandafter\ifx\csname natexlab\endcsname\relax\def\natexlab#1{#1}\fi
\expandafter\ifx\csname bibnamefont\endcsname\relax
  \def\bibnamefont#1{#1}\fi
\expandafter\ifx\csname bibfnamefont\endcsname\relax
  \def\bibfnamefont#1{#1}\fi
\expandafter\ifx\csname citenamefont\endcsname\relax
  \def\citenamefont#1{#1}\fi
\expandafter\ifx\csname url\endcsname\relax
  \def\url#1{\texttt{#1}}\fi
\expandafter\ifx\csname urlprefix\endcsname\relax\def\urlprefix{URL }\fi
\providecommand{\bibinfo}[2]{#2}
\providecommand{\eprint}[2][]{\url{#2}}

\bibitem[{\citenamefont{Mislo}(2002)}]{Mislow02}
\bibinfo{author}{\bibfnamefont{K.}~\bibnamefont{Mislo}},
  \emph{\bibinfo{title}{Introduction to Stereochemistry}}
  (\bibinfo{publisher}{Dover}, \bibinfo{address}{Mineola, New York},
  \bibinfo{year}{2002}).

\bibitem[{\citenamefont{Johnson}(2005)}]{Johnson05}
\bibinfo{author}{\bibfnamefont{L.~N.} \bibnamefont{Johnson}},
  \bibinfo{journal}{European Review} \textbf{\bibinfo{volume}{13}},
  \bibinfo{pages}{77} (\bibinfo{year}{2005}).

\bibitem[{\citenamefont{Kral et~al.}(2003)\citenamefont{Kral, Thanopulos,
  Shapiro, and Cohen}}]{Kral03}
\bibinfo{author}{\bibfnamefont{P.}~\bibnamefont{Kral}},
  \bibinfo{author}{\bibfnamefont{I.}~\bibnamefont{Thanopulos}},
  \bibinfo{author}{\bibfnamefont{M.}~\bibnamefont{Shapiro}}, \bibnamefont{and}
  \bibinfo{author}{\bibfnamefont{D.}~\bibnamefont{Cohen}},
  \bibinfo{journal}{Physical Review Letters} \textbf{\bibinfo{volume}{90}},
  \bibinfo{pages}{033001} (\bibinfo{year}{2003}).

\bibitem[{\citenamefont{Kral et~al.}(2005)\citenamefont{Kral, Thanopulos, and
  Shapiro}}]{Kral05}
\bibinfo{author}{\bibfnamefont{P.}~\bibnamefont{Kral}},
  \bibinfo{author}{\bibfnamefont{I.}~\bibnamefont{Thanopulos}},
  \bibnamefont{and} \bibinfo{author}{\bibfnamefont{M.}~\bibnamefont{Shapiro}},
  \bibinfo{journal}{Physical Review A} \textbf{\bibinfo{volume}{72}},
  \bibinfo{pages}{020303} (\bibinfo{year}{2005}).

\bibitem[{\citenamefont{Gerbasi et~al.}(2006)\citenamefont{Gerbasi, Shapiro,
  and Brumer}}]{Gerbasi06}
\bibinfo{author}{\bibfnamefont{D.}~\bibnamefont{Gerbasi}},
  \bibinfo{author}{\bibfnamefont{M.}~\bibnamefont{Shapiro}}, \bibnamefont{and}
  \bibinfo{author}{\bibfnamefont{P.}~\bibnamefont{Brumer}},
  \bibinfo{journal}{Journal of Chemical Physics}
  \textbf{\bibinfo{volume}{124}}, \bibinfo{pages}{074315}
  (\bibinfo{year}{2006}).

\bibitem[{\citenamefont{Hirota}(2012)}]{Hirota12}
\bibinfo{author}{\bibfnamefont{E.}~\bibnamefont{Hirota}},
  \bibinfo{journal}{Proceedings of the Japan Academy Series B-Physical and
  Biological Sciences} \textbf{\bibinfo{volume}{88}}, \bibinfo{pages}{120}
  (\bibinfo{year}{2012}).

\bibitem[{\citenamefont{Patterson et~al.}(2013)\citenamefont{Patterson,
  Schnell, and Doyle}}]{Patterson13}
\bibinfo{author}{\bibfnamefont{D.}~\bibnamefont{Patterson}},
  \bibinfo{author}{\bibfnamefont{M.}~\bibnamefont{Schnell}}, \bibnamefont{and}
  \bibinfo{author}{\bibfnamefont{J.~M.} \bibnamefont{Doyle}},
  \bibinfo{journal}{Nature} \textbf{\bibinfo{volume}{497}},
  \bibinfo{pages}{475} (\bibinfo{year}{2013}).

\bibitem[{\citenamefont{Patterson and Doyle}(2013)}]{Patterson13b}
\bibinfo{author}{\bibfnamefont{D.}~\bibnamefont{Patterson}} \bibnamefont{and}
  \bibinfo{author}{\bibfnamefont{J.~M.} \bibnamefont{Doyle}},
  \bibinfo{journal}{Physical Review Letters} \textbf{\bibinfo{volume}{111}},
  \bibinfo{pages}{023008} (\bibinfo{year}{2013}).

\bibitem[{\citenamefont{Shirley}(1965)}]{Shirley65}
\bibinfo{author}{\bibfnamefont{J.~H.} \bibnamefont{Shirley}},
  \bibinfo{journal}{Phys. Rev.} \textbf{\bibinfo{volume}{138}},
  \bibinfo{pages}{B979} (\bibinfo{year}{1965}).

\bibitem[{\citenamefont{Zare}(1988)}]{Zare}
\bibinfo{author}{\bibfnamefont{R.~N.} \bibnamefont{Zare}},
  \emph{\bibinfo{title}{Angular Momentum: Understanding Spatial Aspects in
  Chemistry and Physics}} (\bibinfo{publisher}{John Wiley \& Sons},
  \bibinfo{address}{Ithaca, NY}, \bibinfo{year}{1988}).

\bibitem[{\citenamefont{Townes and Schawlow}(1955)}]{Townes55}
\bibinfo{author}{\bibfnamefont{C.~H.} \bibnamefont{Townes}} \bibnamefont{and}
  \bibinfo{author}{\bibfnamefont{A.~L.} \bibnamefont{Schawlow}},
  \emph{\bibinfo{title}{Microwave Spectroscopy}}
  (\bibinfo{publisher}{McGraw-Hill}, \bibinfo{address}{New York},
  \bibinfo{year}{1955}).

\end{thebibliography}

\end{document}